# Ultra-compact injection terahertz laser using the resonant inter-layer radiative transitions in multi-graphene-layer structure

ALEXANDER A. DUBINOV,[1,2,3,*] ANDREY BYLINKIN,[3]
VLADIMIR YA. ALESHKIN,[1,2] VICTOR RYZHII,[4,5,6] TAIICHI OTSUJI[4]
AND DMITRY SVINTSOV,[3]

[1]*Institute for Physics of Microstructures of Russian Academy of Sciences, Nizhny Novgorod 603950, Russia*
[2]*Lobachevsky State University of Nizhny Novgorod, Nizhny Novgorod 603950, Russia*
[3]*Laboratory of 2d materials' optoelectronics, Moscow Institute of Physics and Technology, Moscow, Russia*
[4]*Research Institute for Electrical Communication, Tohoku University, Sendai 980-8577, Japan*
[5]*Institute of Ultra High Frequency Semiconductor Electronics of Russian Academy of Sciences, Moscow, 117105, Russia*
[6]*Center for Photonics and Infrared Engineering, Bauman Moscow State Technical University, Moscow 105005, Russia*
*\*sanya@ipmras.ru*

**Abstract:** The optimization of laser resonators represents a crucial issue for the design of terahertz semiconductor lasers with high gain and low absorption loss. In this paper, we put forward and optimize the surface plasmonic metal waveguide geometry for the recently proposed terahertz injection laser based on resonant radiative transitions between tunnel-coupled grapheme layers. We find an optimal number of active graphene layer pairs corresponding to the maximum net modal gain. The maximum gain increases with frequency and can be as large as ~ 500 cm$^{-1}$ at 8 THz, while the threshold length of laser resonator can be as small as ~ 50 μm. Our findings substantiate the possibility of ultra-compact voltage-tunable graphene-based lasers operating at room temperature.

## 1. Introduction

The compact tunable sources of terahertz radiation are highly demanded in security, medical, and telecommunication applications [1–3]. Quantum cascade lasers (QCLs) are considered among most promising candidates to bridge the terahertz gap [4–7], but their operation is currently limited to the cryogenic temperatures only. The reason for low-temperature QCL operation is the strong direct leakage current via closely located subbands in the 1-2 THz range, and phonon-assisted depopulation of the upper subbands in the 4-6 THz range [8]. Also, in the frequency region of 5-12 THz, the operation of QCLs is greatly hindered by the optical phonon absorption (restrallen band) in GaAs and InP [5]. The lead salt diode lasers can be considered as alternative to QCLs as they cover the very broad frequency region (6.5 - 26 THz) [9, 10], but their figures of merit, in particular, output power, are limited due to growth technology problems.

Therefore, the problem of creating the room-temperature terahertz semiconductor laser remains unresolved. In the past ten years, a considerable effort was focused on the creation of graphene-based terahertz lasers with optical and electrical pumping [11–15]. The gapless electron-hole spectrum in graphene allows one to achieve optical gain under population inversion in the whole 1-5 THz range [16]. High optical phonon energy and suppression of the Auger process by the dynamic screening create favorable conditions for relatively long non-radiative recombination lifetimes [17, 18] in graphene. This has enabled the recent demonstration of THz lasing in electrically pumped graphene at 5 THz at liquid nitrogen temperature [19]. However, the prospects of the room-temperature operation of such lasers remain unclear due to the enhancement of recombination processes and self-absorption [20].

Both problems of Drude absorption and interband non-radiative recombination are avoided in the recently proposed lasers based on interlayer electron transitions in tunnel-coupled grapheme layers (GLs) [21–23]. The population inversion between the layers in such structures is achieved by an application of the bias voltage, while the emission of THz photons is due to the photon-assisted interlayer tunneling, similar to that in QCLs (see Fig. 1B). Recent experimental observations of spontaneous photon emission and resonant-tunneling THz detection in the tunnel-coupled GLs [26] motivate further studies on the possibility of full-scale THz lasing.

In this paper, we study the problem of modal gain enhancement in THz lasers based on the resonant photon-assisted tunneling between the GLs. A seemingly simple solution to increase the number of active graphene layer pairs, as we show, does not always lead to the enhancement of modal gain. The reason is that THz gain due to photon-assisted tunneling associated with the transverse component of electric field always competes with interband and Drude absorption in GLs aided by the lateral electric fields. A similar problem of the waveguide optimization aimed at the absorption reduction is well-known in the design of QCLs [24]. Actually, there exists a limited physical space inside the laser resonator, where the transverse electric field significantly exceeds the lateral field, so that the tunneling gain dominates over the absorption loss. In this paper, we find the corresponding upper limit of the modal gain in the THz lasers based on the multiple tunnel-coupled GLs. To this end, we combine the microscopic model of photon-aided resonant tunneling in graphene double layers with electromagnetic simulations of the TM-mode surface plasmon waveguides. Depending on the lasing frequency, the ultimate gain can reach from ~ 50 cm$^{-1}$ (at 4 THz) to ~ 500 cm$^{-1}$ (at 10 THz) and is achieved for the waveguide structures with dozens of active graphene double layers. The corresponding threshold length can be as low as 40 - 300 μm, depending on the frequency of operation. Our findings support the possibility of creating ultra-compact graphene-based THz lasers.

## 2. Device model

We primarily consider the device structure shown in Fig. 1A. It consists of *N* pairs of independently-contacted GLs separated by a narrow tunneling barrier layer of thickness *d* (2.5 nanometers of WS$_2$). To increase the density of tunneling carriers, we assume the layers to be doped by donors (n-GL, marked with red in Fig. 1A) and acceptors (p-GL, marked with blue), respectively. Each pair of the active GLs is separated by a narrow insulating layer of thickness *f* (90 nanometers of hexagonal boron nitride, hBN). The doping with density of $10^{12}$ cm$^{-2}$ leads to the formation of the two-dimensional electron (2DEG) and the two-dimensional hole (2DHG) gases in the n-GL (the top GL of each pair) and the p-GL (the bottom GL of each pair), respectively. In the structure under consideration, the height of the contacts *H* is related to the number of double graphene layers *N* and the distance *f* between them via $H = N \times (f + d)$.

The bias voltage *V* between the side contacts induces the extra charges of opposite sign in the layers and the electric field between them, which causes the inter-GL population inversion (Fig. 1 B). Under the population-inverted conditions, the stimulated electron tunneling accompanied by the photon emission is more probable than the inverse absorptive process. Thus, such a GL structure can serve as the laser gain medium with the lateral injection pumping. The active graphene structure is sandwiched by the hBN insulator cladding layers with total thickness *W* = 10 μm equal to half wavelength of THz photon of frequency $\omega/2\pi$ = 6 THz in the insulator. The whole structure is placed a top a metal substrate constituting a surface plasmon waveguide for the TM-modes propagating in the *y*-direction in Fig. 1A. The electrical contacts are separated by an insulating gap with thickness *w* = 90 nm from the waveguiding surface.

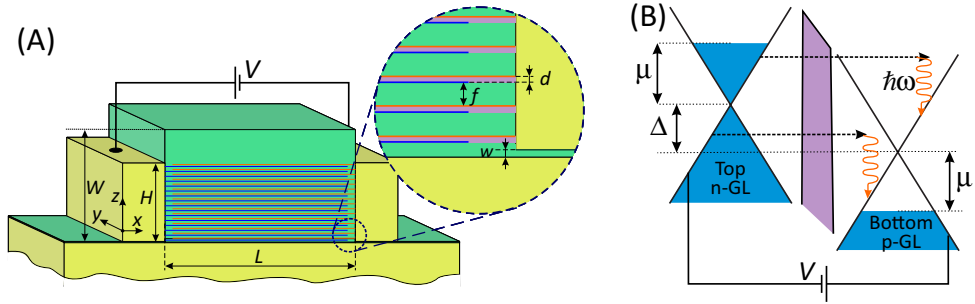

Fig. 1. (A) Schematic view of the terahertz laser with a stack of tunnel-coupled graphene layers (N pairs) with side injection embedded in a surface plasmon waveguide (B) Band diagram illustrating the process of stimulated photon-assisted resonant tunneling under application of interlayer voltage V

The net modal gain g(ω) for the TM-mode in the device under consideration is given by the following expression:

$$g(\omega) = -\frac{4\pi n_{eff}(\omega, N)}{c\kappa}\left[\operatorname{Re}\sigma_{zz}(\omega)\Gamma_z(\omega) + \operatorname{Re}\sigma_{yy}(\omega)\Gamma_y(\omega)\right], \qquad (1)$$

where

$$\Gamma_{z,y}(\omega) = \frac{\sum_{j=0}^{N-1} \int_0^L |E_{z,y}(x, w+(f+d)j, \omega)|^2 dx}{\int_{-\infty}^{+\infty}\int_{-\infty}^{+\infty} |E_z(x,z,\omega)|^2 dxdz} \qquad (2)$$

are the gain-overlap factors, $n_{eff}(\omega, N)$ is the effective refractive index of the mode, $E_{z,y}(x, z, \omega)$ are the components of the electric field in TM-mode, $c$ is the light speed in vacuum, $L$ is the waveguide width, $\kappa$ is the dielectric constant of the barrier layer, $\sigma_{zz}(\omega)$ and $\sigma_{yy}(\omega)$ are the transverse and lateral dynamic conductivities of tunnel-coupled graphene double layer structure.

The inter-GL transitions assisted by the emission of TM-photons with the dominant z component of electric field conserve the in-plane electron momentum and, hence, do not involve scattering (resonant-tunneling photon-assisted transitions). The resonant character of tunneling transitions results in the Lorentzian shape of the real part of tunnel conductivity [25]

$$\text{Re}\,\sigma_{zz}(\omega) \cong -2\frac{e^2}{\hbar}|z_{u,l}|^2 \Sigma \frac{\omega\gamma}{(\omega-\omega_{max})^2 + \gamma^2}, \qquad (3)$$

where $\Sigma$ is the density of electrons (holes) in the pertinent GLs, $ez_{u,l}$ is the dipole matrix element between wave functions of carriers localized on the upper and lower layers, and $\gamma$ is the relaxation broadening of the tunnel resonance which is found to be ~ 3 meV for the given carrier density (see Appendix for the detailed calculation). As seen from the band diagram in Fig. 1B, the maximum conductivity corresponds to the transitions with emission of photon with energy $\hbar\omega_{max}$ equal to the spacing between Dirac points in the neighboring GLs Δ. This energy should corrected by the value of depolarization shift [29]

$$\Delta_{dep} = -8\pi \frac{e^2 |z_{u,l}|^2 \Sigma}{\kappa d} \qquad (4)$$

where $\kappa$ and $d$ are the permittivity and thickness of the tunneling barrier layer. The resonantconductivity at $\omega = \omega_{max}$ reads as follows

$$\text{Re}\,\sigma_{zz}(\omega_{max}) \cong -2\frac{e^2}{\hbar}|z_{u,l}|^2 \Sigma \frac{\omega_{max}}{\gamma} \qquad (5)$$

The dipole matrix element $z_{u,l}$ is evaluated as

$$z_{u,l} = \int_{-\infty}^{\infty} \varphi_u^*(z) z \varphi_l(z) dz, \qquad (6)$$

where $\varphi_u(z)$ and $\varphi_l(z)$ are the z-dependent envelope wave functions of electrons located at the upper and lower GLs, respectively. These wave functions are found by solving the effective Schrodinger equation where each graphene layer is modeled as a one-dimensional delta-well [23]. The strength of the delta-well is chosen to provide the correct work function from graphene to the surrounding dielectric (0.4 eV for graphene embedded in $WS_2$ [27]). The effective mass of electron in $WS_2$ is taken to be 0.27 of the free electron mass [28].

The in-plane conductivity of graphene layers is given by [30]

$$\mathrm{Re}\,\sigma_{yy}(\omega) = \frac{e^2}{\hbar}\left[\frac{2\gamma\mu}{\pi(\hbar^2\omega^2+\gamma^2)} + \frac{1}{4}\{f(-\hbar\omega/2) - f(\hbar\omega/2)\}\right], \quad (7)$$

where the first term in curly brackets is responsible for Drude absorption and the second one – for the interband absorption, $\mu$ is the Fermi energy of carriers (electrons in top GL and holes in bottom one), and $f(\varepsilon) = \left[1 + e^{(\varepsilon-\mu)/kT}\right]^{-1}$ is the Fermi function.

In the case of relatively short GLs (for the diffusion length of electrons and holes $L_d \gg L$), the carrier Fermi Fermi energy $\mu$ is constant throughout the layer and related to the applied voltage $V$ and spacing between the Dirac points $\Delta$ as $\Delta = eV - 2\mu$. At strong doping and/or low temperatures such that $\mu \gg k_BT$, this Fermi energy is a square-root function of density, $\mu = \hbar v_W \sqrt{\pi\Sigma}$, where $v_W \approx 10^8$ cm/s is the characteristic velocity of electrons and holes in GLs. This yields the following relation between $\Delta$ and $V$:

$$\Delta = e\left(V + V_0 - \sqrt{V_0^2 + V_t^2 + 2VV_0}\right), \quad (8)$$

where $eV_0 = \dfrac{\hbar^2 v_W^2 \kappa}{2e^2 d}$ and $V_t = 2\hbar v_W \sqrt{\pi\Sigma_i}/e$ is the threshold voltage. At $V \le V_t$, Eq. (8) yields $\Delta \le 0$, while at $V > V_t$, $\Delta > 0$. In latest case the photon emission upon interlayer intraband tunneling dominates over the absorption.

The solution of the Maxwell equations with the pertinent complex permittivity of the surface plasmon waveguide and metal strips yields the spatial distributions of the electric fields, $E_{z,y}(x,z,\omega)$ in the propagating mode and consequently, the gain-overlap factors. These equations were solved numerically using 2D mode analysis in COMSOL. It is assumed that the metal strips and substrate in the surface plasmon waveguide are made of Au [31]. The h-BN complex permittivity was extracted from Ref. [32]. In the calculations we have neglected the influence of GLs on the field distributions, because the electric field component along the GLs and, hence, the surface currents are small in the TM mode, as we shall discuss below.

### 3. Results and discussion

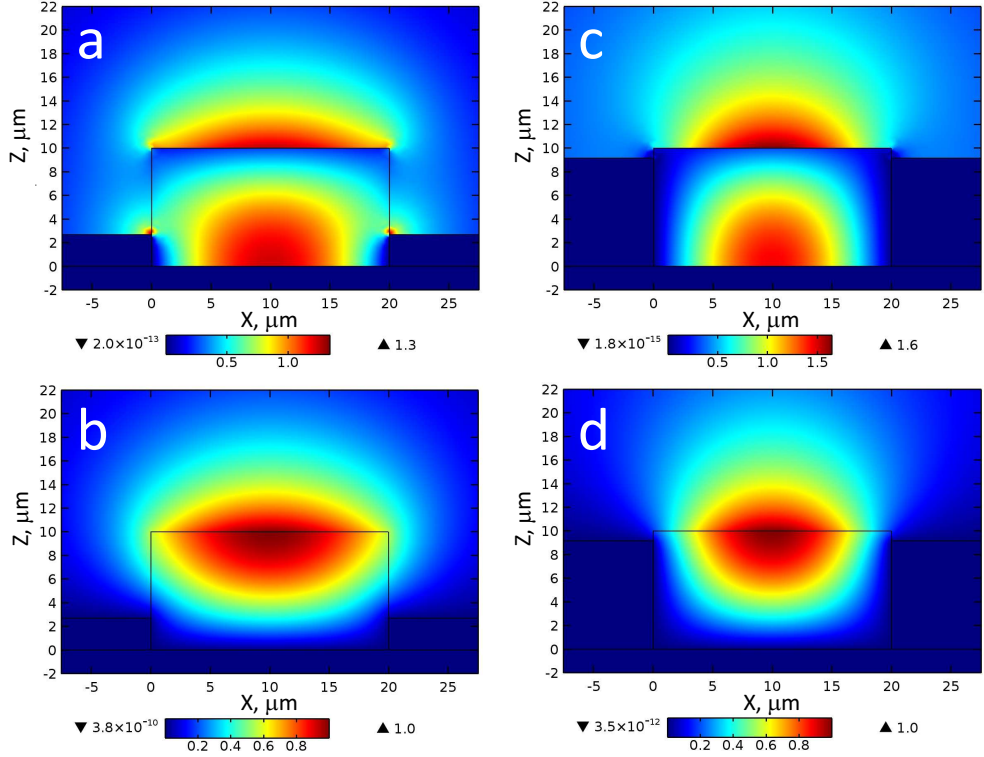

Fig. 2. Spatial distributions of the photon electric field components in TM mode of the surface plasmonic waveguide under consideration for two different heights of the metal contacts, corresponding to $N$ = 30 (a,b) and $N$ = 100 (c,d). Panels (a, c) show the $z$-component of electric field, panels (b, d) show the $y$-component. The wave frequency is $\omega/2\pi$ = 5 THz.

Figures 2 (a)-2 (d) show the examples of the spatial distributions of the photon electric field components $|E_z(x,z,\omega)|$ and $|E_y(x,z,\omega)|$ in the TM mode of a surface plasmon waveguide in device from Fig. 1. The obtained dependences correspond to $\omega/2\pi$ = 5 THz. In the waveguide under consideration, the vertical component of electric field induced by the interlayer tunneling is dominantly localized near the bottom metal surface. The lateral component of electric is, on the contrary, localized at the air/dielectric interface. This creates favorable conditions for the tunneling gain if the graphene double-layers are localized close to the bottom metal surface.

Increasing the number of active graphene layers and, hence, the height of metal contacts, one increases the gain due to interlayer tunneling transitions. On the other hand, the topmost layers are located closer to the maximum of in-plane field $E_y(x,z,\omega)$, as seen in Fig. 2, and increase in the layer number can significantly increase ohmic losses (due to both Drude and interband absorption). Thus, there should exist an optimal number of active layers for which the net modal gain reaches its maximum.

Fig. 3A shows the frequency (for $\omega = \omega_{\max}$) dependences of the modal gain calculated with Eq. (1) for several fixed numbers of active double layer structures. From Fig. 3A one can see that the modal gain is higher for structures with a small number of GL pairs in the low-frequency region. For low frequencies Drude absorption increases faster than the gain with an increase in the number of GL pairs. The situation is reversed for the high frequency region

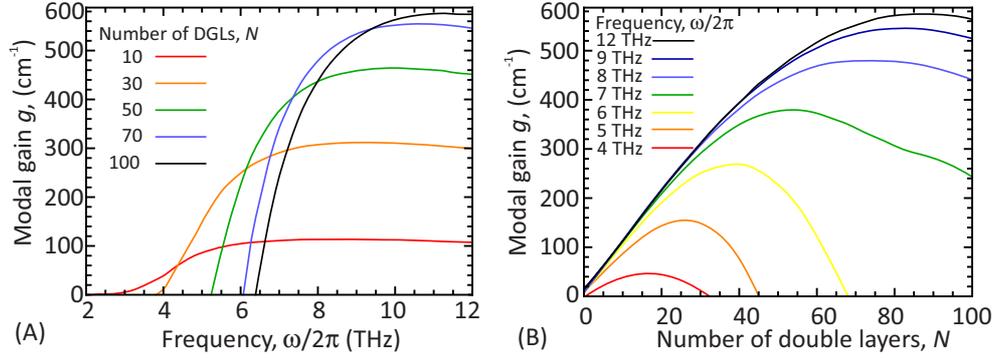

Fig. 3. (A) Frequency dependence of modal gain at fixed numbers of double graphene layers *N* (B) Dependence of modal gain on the number of double layers at fixed frequencies. The increase in *N* above the threshold value leads to increase in Drude absorption and reduction in modal gain.

The gain threshold, i.e. the frequency for which the gain exceeds the loss in the GLs, shifts to the higher frequencies with increasing *N*.

The presence of the optimal number of active double layers $N_{opt}$ is seen from Fig. 3B, where plot the dependence of modal gain on *N* at several fixed frequencies corresponding to the interlayer tunneling resonance. The modal gain maximum increases and shifts toward higher values of *N* with increasing frequency. For example, for $\omega/2\pi$ = 5 THz $N_{opt}$ ~ 27, and for $\omega/2\pi$ = 8 THz $N_{opt}$ ~ 72. Furthermore, with increasing frequency the range of the number of GL pairs at which the gain is possible becomes broader.

Fig. 4 shows the frequency dependences of $N_{opt}$ and the corresponding maximum modal gain. For structure considered in our work, the maximum modal gain becomes positive at $\omega/2\pi$ = 3 THz ($N_{opt}$ ~ 18) and increases up to 580 cm$^{-1}$ with increasing the frequency. The increase in maximum modal gain at moderate frequencies ($\omega$ < 10 THz) is due to the reduction in Drude absorption and increase in the resonant tunnel conductivity [Eq. (5)] due to the factor $\omega_{max}$ in the numerator. At higher frequencies, the maximum attainable gain saturates because large resonant frequencies require the application of large interlayer voltage which, in turn, reduces the dipole matrix elements $ez_{u,l}$. This dependence of maximum tunnel conductivity is illustrated in the inset of Fig. 4, where we plot the dimensionless 'gain factor' $4\pi\sigma_{zz}(\omega_{max})/c\sqrt{\kappa}$.

Using Eq. (1) one can determine the threshold length $L_{th}(\omega)$ of the structure for which the lasing condition is fulfilled: .

$$L_{th}(\omega) = \frac{\ln[1/R(\omega)]}{g(\omega) - \alpha(\omega)}, \qquad (9)$$

where $\alpha(\omega)$ is the surface plasmon waveguide absorption coefficient due to losses in metal, and $R(\omega)$ is the reflection coefficient on structure edges. If the resonator mirrors represent the cleaved surfaces of h-BN, $R(\omega)$ can be estimated as follows:

$$R(\omega) = \left(\frac{n_{eff}(\omega) - 1}{n_{eff}(\omega) + 1}\right)^2 \qquad (10)$$

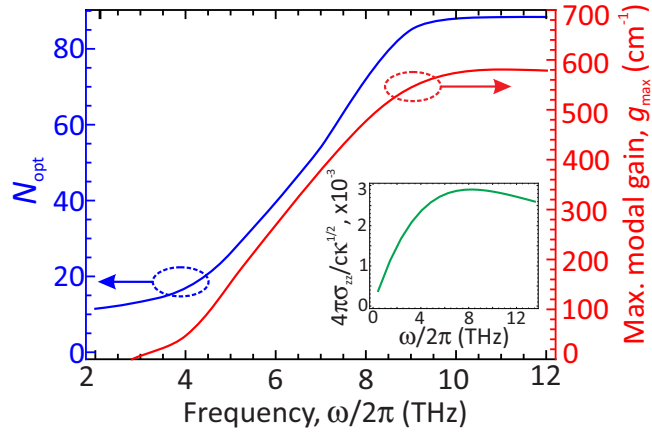

Fig. 4. Frequency dependence of the optimal number of double graphene layers $N_{opt}$ and the corresponding maximum modal gain. Inset: frequency dependence of dimensionless 'gain factor' $4\pi\sigma_{zz}(\omega_{max})/c\sqrt{\kappa}$

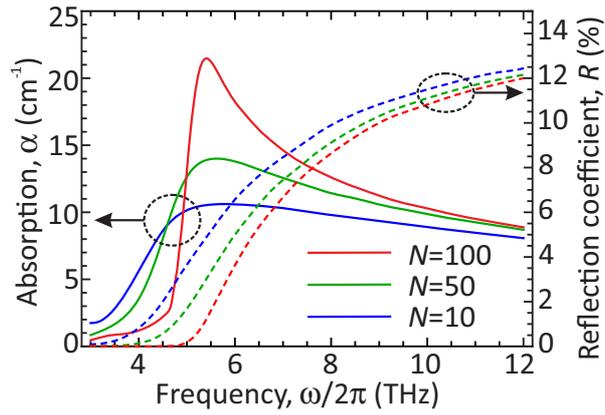

Fig. 5. Frequency dependence of waveguide absorption and reflection coefficients different values of $N$

Fig. 5 shows the frequency dependence of the absorption coefficient $\alpha(\omega)$ and the reflection coefficient $R(\omega)$ in the structure under consideration for three different values of $N$. From Fig. 5 one can see that reflection coefficients increase with increasing the frequency for our waveguide structure. At the same time, a fairly weak dependence of reflection coefficients on $N$ is observed. The absorption coefficients have maxima at 5.5 THz, and the frequency corresponding to the maximum absorption depends strongly on the waveguide geometry. Below 5.5 THz, the electromagnetic wave is weakly bound to the waveguide, which leads to the low Ohmic losses in metal. The absorption maximum corresponds to the wave with strong electric field at the side contacts. At higher frequencies, the electric field of the wave is localized primarily at the bottom metal surface, aside from the waveguide sidewalls, which again leads to the reduction in absorption.

Fig. 6 shows the frequency dependence of laser threshold length $L_{th}(\omega)$ for different values of $N$ in the given structure. The trends observed in Fig. 6 are similar to those in Fig. 3. For each frequency, there exists an optimal value of $N$. In the frequency range of interest (5-12 THz, where the operation of quantum cascade lasers is strongly hindered by the semiconductor restrallen bands, $L_{th}(\omega)$ can be reduced down to values of 40 - 300 μm by a proper selection of *N*. Such laser threshold lengths are considerably less than the characteristic lengths of quantum cascade lasers (over 1000 μm [34]). Furthermore, this means that one can use a sufficiently small area graphene layers for laser structures (20 × 40 μm) which typically demonstrate better electronic properties.

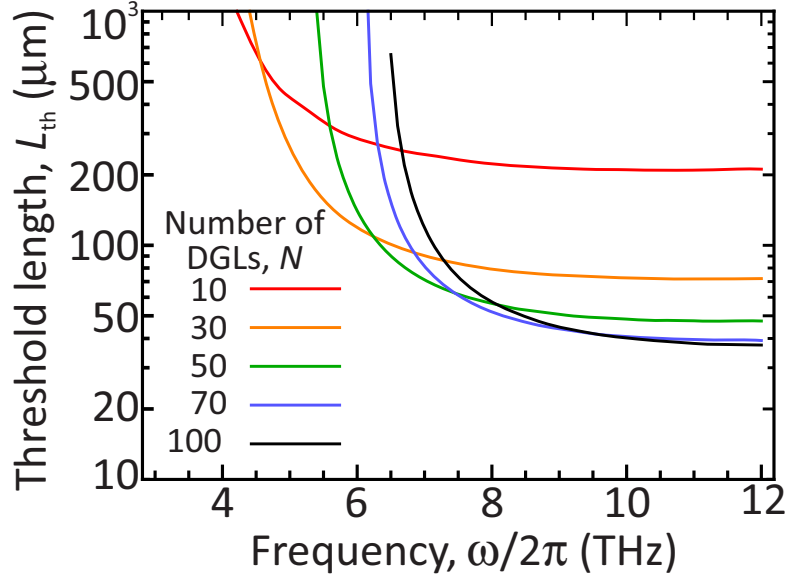

Fig. 6. Frequency dependence of laser threshold length for different values number of double grapheme layers (DGLs) *N*

## 4. Conclusion

We have proposed and substantiated the concept of injection THz laser based on the multiple graphene-layer structure exploiting the interlayer resonant radiative transitions and embedded in plasmonic waveguide. An increase in the number of active pairs of tunnel-coupled grapheme layers can lead to a dramatic increase in the modal gain (up to ~500 cm$^{-1}$) as compared with quantum cascade lasers. We have found, however, that there exists an optimal number of active double graphene layers for each frequency. Further increase of layer number would lead to an increase in interband and Drude absorption in the structure. The graphene-based lasers with the surface plasmon waveguides considered above have advantages associated with an efficient voltage tuning, low Drude absorption and optical phonon absorption. This would enable their operation in the frequency range from 5 to 12 THz.

## Acknowledgments

The work was supported by Grant No. 16-19-10557 of the Russian Science Foundation.

## Appendix: Calculation of the tunnel resonance broadening

The broadening factor $\gamma$ entering the expressions for tunnel conductivity is different from the collisional broadening of electron energy in a single layer and from the quantity $\hbar\tau_{tr}^{-1}$, where $\tau_{tr}$ is the transport relaxation time. The origin of this difference is the interference of carrier scattering events in the tunnel-coupled layers: if the action of impurity on carriers in both layers were the same, there would be no scattering and, hence, no interlayer current [25,33]. The account for this interference leads to the following expression for $\gamma$

$$\gamma = 2\pi \sum_{\mathbf{q}} \left\langle |V_t(\mathbf{q}) - V_b(\mathbf{q})|^2 \right\rangle \delta[v_W(p - |\mathbf{p}-\mathbf{q}|)](1 - \cos\Theta_{\mathbf{p},\mathbf{p}-\mathbf{q}})/2 \qquad (11)$$

where $V_t(\mathbf{q})$ and $V_b(\mathbf{q})$ are the Fourier components of impurity potential acting on the electrons in bottom and top layers, the factor $(1-\cos\Theta_{\mathbf{p},\mathbf{p}-\mathbf{q}})/2$ comes from the overlap of chiral wave functions in graphene, and the brackets denote averaging over random impurity positions. Performing this averaging, we find

$$\gamma = 2\pi(\Sigma_{i,t} + \Sigma_{i,b}) \sum_{\mathbf{q}} V_0^2(\mathbf{q})(1 - e^{-qd})^2 \delta[v_W(p - |\mathbf{p}-\mathbf{q}|)](1 - \cos\Theta_{\mathbf{p},\mathbf{p}-\mathbf{q}})/2 \qquad (12)$$

where $\Sigma_{i,t}$ and $\Sigma_{i,b}$ are the impurity densities in top and bottom layers, $V_0(\mathbf{q}) = 2\pi e^2 / \kappa(q + q_{TF})$ is the Fourier transform of the Coulomb potential created by a single impurity, and $q_{TF}$ is the Thomas-Fermi screening wave vector. The small factor $1 - e^{-qd} \ll 1$ is due to the discussed interference of scattering events. The evaluation of Eq. (12) with the following parameters $\Sigma_{i,t} = \Sigma_{i,b} = 10^{12}$ cm$^{-2}$, $\kappa = 5$, $d = 2.5$ nm, carrier energy $pv_W = 110$ meV (corresponding to the Fermi energy at given density) leads us to $\gamma \approx 3$ meV.